\documentclass[]{pasj01}
\Received{}%{yyyy/mm/dd}
\Accepted{}%{yyyy/mm/dd}
\usepackage{bm}
\usepackage{comment}
\usepackage[switch,mathlines]{lineno}
\usepackage{color}
\newcommand{\rev}[1]{\textcolor{black}{#1}}
\begin{document} 

\title{ 
%\LETTERLABEL %%% <-- uncomment for LETTER article  
%\REVIEWLABEL %%% <-- uncomment for REVIEW article  
Nodal precession of a hot Jupiter transiting the edge of a late A-type star TOI-1518}

%%% begin: list of authors
% Do NOT capitalize all letters in "textsc".
\author{Noriharu \textsc{Watanabe}\altaffilmark{1}}
\email{n-watanabe@g.ecc.u-tokyo.ac.jp}

\author{Norio \textsc{Narita},\altaffilmark{2,3,4}}
\email{narita@g.ecc.u-tokyo.ac.jp}

\author{Yasunori \textsc{Hori}\altaffilmark{3,5}}
\email{yasunori.hori@nao.ac.jp}

\altaffiltext{1}{Department of Multi-Disciplinary Sciences, Graduate School of Arts and Sciences, The University of Tokyo, 3-8-1 Komaba, Meguro, Tokyo 153-8902, Japan}
\altaffiltext{2}{Komaba Institute for Science, The University of Tokyo, 3-8-1 Komaba, Meguro, Tokyo 153-8902, Japan}
\altaffiltext{3}{Astrobiology Center, 2-21-1 Osawa, Mitaka, Tokyo 181-8588, Japan}
\altaffiltext{4}{Instituto de Astrof\'{i}sica de Canarias (IAC), 38205 La Laguna, Tenerife, Spain}
\altaffiltext{5}{National Astronomical Observatory of Japan, 2-21-1 Osawa, Mitaka, Tokyo 181-8588, Japan}
%\altaffiltext{6}{Departamento de Astrof\'isica, Universidad de La Laguna (ULL), E-38206 La Laguna, Tenerife, Spain}


\maketitle

\begin{abstract}
TOI-1518b, a hot Jupiter around a late A-type star, is one of the few planetary systems that transit the edge of the stellar surface (the impact parameter $b\sim0.9 $) among hot Jupiters around hot stars (Cabot et al. 2021). The high rotation speed of the host star ($\sim85$ km s$^{-1}$) and the nearly polar orbit of the planet ($\sim \timeform{120D}$) may cause a nodal precession.
In this study, we report the nodal precession undergone by TOI-1518\,b. This system is the fourth planetary system in which nodal precession is detected.
We investigate the time change in $b$ from the photometric data of TOI-1518 acquired in 2019 and 2022 with TESS and from the spectral transit data of TOI-1518b obtained in 2020 with two high-dispersion spectrographs; CARMENES and EXPRES.
We find that the value of $b$ is decreasing with $db/dt=-0.0116\pm0.0036$\,year$^{-1}$, indicating that the transit trajectory is moving toward the center of the stellar surface.
We also estimate the minimum value of the quadrupole mass moment of TOI-1518 $J_{2,\mathrm{min}}=4.41\times 10^{-5}$ and the logarithm of the Love number of TOI-1518 $\log{k_2}= -2.17\pm 0.33$ from the nodal precession.

\end{abstract}
%\linenumbers

\section{Introduction}
To date, 20 hot Jupiters have been discovered around hot stars whose effective temperatures are above 7,000K.
These hot stars have a wide range of obliquities, that is, angles between the stellar rotational and orbital axes.
The observed spin-orbit misalignment trends of hot Jupiters around hot stars imply that they did not undergo tidal realignment because of their shallow convective envelopes \citep{2012ApJ...757...18A}. 
Hot stars barely sustain stellar winds that lose their spin angular momentum due to magnetic braking.
Hot stars tend to rotate rapidly as is known for the Kraft break \citep{1967ApJ...150..551K}.
The oblateness of fast-rotating stars causes nodal precession of hot Jupiters in misaligned orbits.
Nodal precession of three hot Jupiters on nearly polar orbits around rapidly rotating hot stars: Kepler-13Ab \citep{2012MNRAS.421L.122S,2014MNRAS.437.1045S,2018AJ....155...13H} and WASP-33b \citep{2015ApJ...810L..23J,2020PASJ...72...19W,2022MNRAS.512.4404W,2022ApJ...931..111S}, and KELT-9b \citep{2022ApJ...931..111S}, were detected.
The nodal precession of a planet enables us to restrict the quadrupole mass moment $J_2$ and the Love number $k_2$. $J_2$ \rev{indicates} the oblateness of the host star \rev{and its internal mass redistribution due to its rapid rotation}. The Love number $k_2$ expresses the rigidity of the internal structure, which could be an important clue for understanding the susceptibility to the tidal effects.

\renewcommand{\thefootnote}{\fnsymbol{footnote}}

\citet{2021AJ....162..218C} have discovered a hot Jupiter (planetary radius: $R_p=1.875\pm 0.053 R_J$, orbital period: $P_{\mathrm{orb}}=1.902603\pm 0.000011$days, scaled semi-major axis: $a/R_s=4.291^{+0.057}_{-0.061}$) around a rapidly-rotating late A-type star TOI-1518 with the effective temperature $T_\mathrm{eff}=7300\pm100$K and the projected rotational speed $V\sin i_s =85.1\pm 6.3\,\mathrm{km\ s^{-1}}$, where $i_s$ is the angle between the stellar spin axis and the line of sight. They have measured the projected spin-orbit obliquity $\lambda=\timeform{-119D66}^{+0.98}_{-0.93}$ 
\footnote{They defined the range of $\lambda$ as $\timeform{0D}<\lambda<\timeform{360D}$, so they measured $\lambda=\timeform{240D34}^{+0.93}_{-0.98}$. In this study, we define this range as $\timeform{-180D}<\lambda<\timeform{180D}$. Thus, we write $\lambda=\timeform{-119D66}^{+0.98}_{-0.93}$.}
and the impact parameter $b=0.9036^{+0.0061}_{-0.0053}$, indicating that the planet transits the edge of the stellar surface in a near-polar orbit. If the orbit of TOI-1518\,b shifts toward a larger b via nodal precession, then the transit of TOI-1518\,b will finish in several decades.

Section 2 presents our measurements of the impact parameter of TOI-1518b from the photometric data of The Transiting Exoplanet Survey Satellite (TESS; \cite{2015JATIS...1a4003R}) and transit spectral data from high-resolution spectrographs. In Section 3, we describe the change of the impact parameter and discuss the nodal precession results. Finally, we present our conclusions in Section 4.

\section{Observations and Analysises}

\begin{figure*}[htbp]
 \begin{center}
  \includegraphics[width=\linewidth]{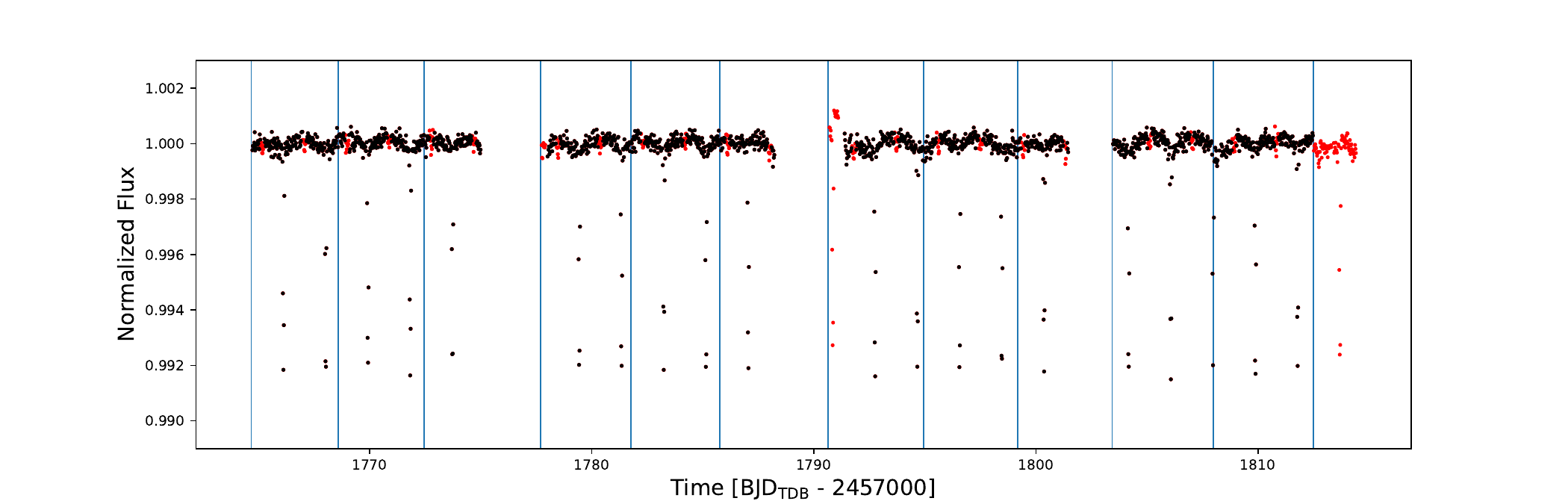} 
 \end{center}
 \begin{center}
  \includegraphics[width=\linewidth]{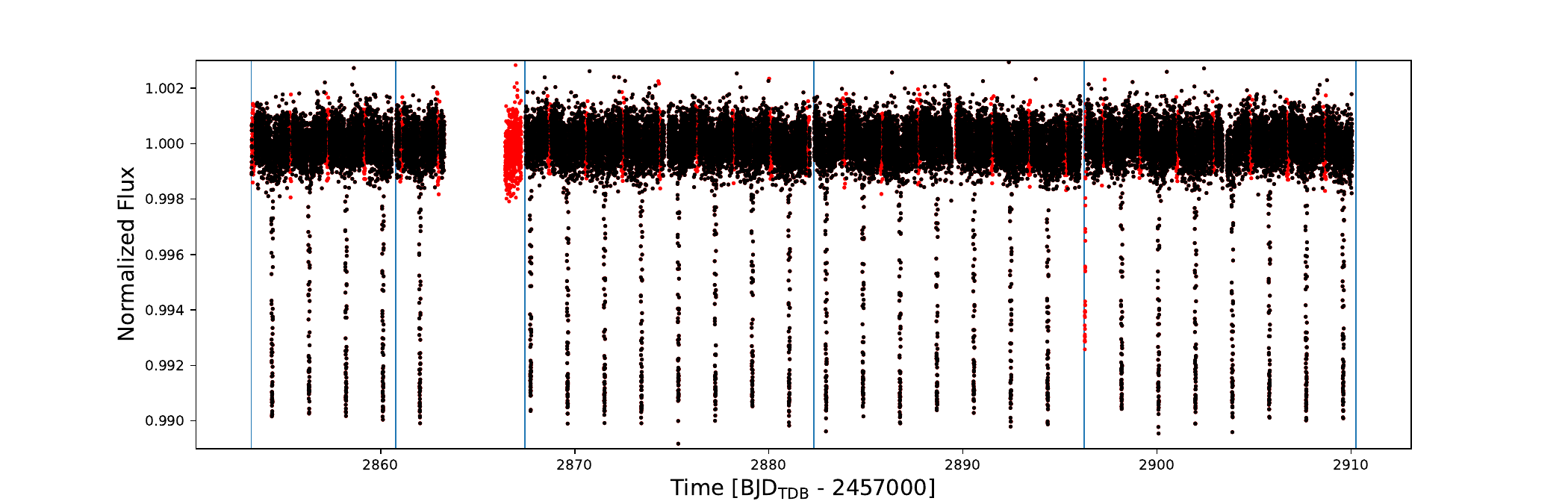} 
 \end{center}
\caption{The normalized light curve of TOI-1518 in 2019 (top figure) and the one in 2022 (bottom figure) from Presearch Data Conditioning SAP (PDCSAP). The vertical blue lines show the scheduled momentum dumps. The red points are the dimming parts caused by the secondary eclipse and the small discontinuities and flux ramps, which are excluded from our analysis.}\label{fig_TESS}
\end{figure*}

\subsection{TESS Photometry}
TESS observed TOI-1518 from UT 2019 October 7 to November 27 (Sectors 17 and 18) and from UT 2022 September 30 to November 26 (Sectors 57 and 58). \citet{2021AJ....162..218C} utilized the TESS-SPOC HLSP light curves \citep{2020RNAAS...4..201C} in 2019. In addition to these datasets, we acquired the datasets of the SPOC light curves \citep{2016SPIE.9913E..3EJ} in 2022 for TOI-1518 from the Mikulski Archive for Space Telescopes (MAST). The exposure times are 30 minutes in Sectors 17 and 18, and 2 minutes in Sectors 57 and 58. We used the Presearch Data Conditioning SAP (PDCSAP) light curves, which were included in the SPOC datasets and corrected for systematic trends using other sources on the TESS detector. Subsequently, we excluded the small discontinuities and flux ramps due to the momentum dumps, and the dimming parts due to the secondary eclipse. Figure \ref{fig_TESS} shows the light curves of TOI-1518 obtained from TESS data.

We created light-curve models of TOI-1518 using \texttt{PyTransit} \citep{Parviainen2015} by a supersampling method to calculate an accurate model of the transit light curve. As the eccentricity is negligible ($e=0.0031^{+0.0047}_{-0.0022}$; \cite{2021AJ....162..218C}), which is represented by a circular orbit. Then we used a Matern 3/2 kernel for the Gaussian process to fit the wavelet-shaped features using \texttt{celerite} \citep{2017AJ....154..220F}. 

We fitted the light curves in two epochs to the models with the following 12 parameters using Markov Chain Monte Carlo (MCMC): impact parameter $b$ in each year, two hyperparameters of Matern 3/2 kernel, the radius ratio $R_{p}/R_{s}$, mid-transit time $T_{0}$, the orbital period $P_{\mathrm{orb}}$, semi-major axis normalized by stellar radii $a/R_{s}$, two quadratic limb darkening coefficients $u_1$ and $u_2$, and photometric jitter term $\sigma_{\mathrm{jit}}$ in each year. 
For this fitting, we set the logarithm of the likelihood, $\ln L_{\mathrm{like}}$ as
\begin{equation}
\label{post_tess}
\ln L_{\mathrm{like}} =-\frac{1}{2}\left(\mathbf{r}^{T}\mathbf{K}_{\mathrm{ker}}^{-1}\mathbf{r}+\ln |\mathbf{K}_{\mathrm{ker}}|\right),
\end{equation}
where $\mathbf{r}$ is a series of residuals obtained by subtracting the model data from the observation data and $\mathbf{K}_{\mathrm{ker}}$ is the kernel whose element is described as follows:
\begin{eqnarray}
\label{kern_tess}
k_{i, j} &=&\alpha^{2} \left( 1+\frac{\sqrt{3}|t_{i}-t_{j}|}{l} \right)\exp\left({-\frac{\sqrt{3}|t_{i}-t_{j}|}{l}}\right)  \nonumber \\
   & &+\left\{ \left(\sigma_{i}^{2} + \sigma_{\mathrm{jit}}^{2} \right) \delta_{i,j}\right\}.
\end{eqnarray}
$t_{i}$ and $t_{j}$ are the observation times, $\sigma_{i}$ is the error of data point $i$ and $\delta_{i,j}$ is the Kronecker delta.
To calculate these parameter values, we ran 10,000 steps, cut off the first 5,000 steps as burn-ins, and iterated this set 32 times.
Table \ref{TESSresult} shows the fitted values and priors for the MCMC fitting, and Figure \ref{MCMC_TESS} in the Appendix shows the posterior distributions.

\begin{table*}[htbp]
  \tbl{Parameters of TOI-1518b from the light curve of TESS. In the column of "Prior", we describe uniform priors as $\mathcal{U}$(lower bound, upper bound) and normal priors as $\mathcal{N}$(mean, standard deviation).}{%
  \begin{tabular}{lcc}
      \hline
      Parameter & Fitted Value & Prior for Fitting \\ 
      \hline
      Impact Parameter in 2019 $b_{2019}$       & $0.91497 ^{+0.00093}_{-0.00089} $          & $\mathcal{U}(0,1)$ \\
      Impact Parameter in 2022 $b_{2022}$     & $0.8797 \pm 0.0011$             & $\mathcal{U}(0,1)$ \\
      Orbital Period (days) $P_{\mathrm{orb}}$ & $1.90261178 ^{+0.00000018}_{-0.00000019}$    & $\mathcal{U}(0,10)$\\
      Mid-transit Time $T_{0}$ (BJD$_{\mathrm{TDB}}$) & $2458766.120581 ^{+0.000095}_{-0.000098} $  & $\mathcal{U}(2458766,2458767)$\footnotemark[$\S$]\\
      Radius Ratio $R_{p}/R_{s}$      & $0.10034 \pm 0.00019$ & $\mathcal{U}(0,0.5)$\\
      Scaled Semi-major Axis $a/R_{s}$          & $4.164 ^{+0.014}_{-0.015}$              & $\mathcal{U}(0,20)$ \\
      Limb-darkening Coefficient $u_{1,\mathrm{TESS}}$         & $0.3323 ^{+0.0016}_{-0.0015}$  & $\mathcal{N}(0.3360,0.0019)$\footnotemark[$*$] \\
      Limb-darkening Coefficient $u_{2,\mathrm{TESS}}$         & $0.1625 ^{+0.0025}_{-0.0026}$  & $\mathcal{N}(0.1587,0.0031)$\footnotemark[$*$] \\
      Log of Hyper Parameter of Amplitude Scale $\ln\sigma$        & $-8.739 ^{+0.058}_{-0.052}$ & $\mathcal{U}(-14,14)$ \\
      Log of Hyper Parameter of Length Scale $\ln\rho$ (days)      & $-1.068 ^{+0.087}_{-0.086}$ & $\mathcal{U}(-7,7)$ \\
      Photometric Jitter Term in 2019 $\sigma_{\mathrm{jit, 2019}}$ (ppm)   & $88.5 ^{+5.6}_{-5.9}$ & $\mathcal{U}(0,10000)$ \\
      Photometric Jitter Term in 2022 $\sigma_{\mathrm{jit, 2022}}$ (ppm)   & $147.2 ^{+7.3}_{-7.7}$ & $\mathcal{U}(0,10000)$ \\
      \hline
    \end{tabular}}\label{TESSresult}
\begin{tabnote}
\footnotemark[$*$] Median and standard deviation of these coefficients are calculated by \texttt{PyLDTk} \citep{Husser2013, Parviainen2015_b} using the values of $T_{\mathrm{eff}}$, $\log g$ and [Fe/H] from \citet{2021AJ....162..218C}.\\ 
\footnotemark[$\S$] This mid-transit time is during the first transit in the TESS observation.
\end{tabnote}
\end{table*}

\subsection{Doppler Tomographic Observation}
We utilized the reduced transit spectral dataset of TOI-1518b from the 3.5m telescope with CARMENES \citep{2014SPIE.9147E..1FQ}, a high-resolution spectrograph ($R\sim 94,600$), at the Calar Alto Observatory on UT 2020 October 8.
These datasets are reduced using the CARACAL pipeline \citep{2014A&A...561A..59Z, 2015A&A...581A.117B} automatically at the end of the observation.
We used a wavelength range from 5180 \AA \ to 7860 \AA \, except for the wavelength regions around the noticeable telluric lines and bad pixels. The dataset contains 14 spectra obtained with an exposure time of 900 s. The signal-to-noise ratio per pixel for each spectrum is $\sim 100$ at 5500 \AA.
We took the continua of these spectra and shifted them to a barycentric frame with \texttt{astropy} \citep{2022ApJ...935..167A} to read these fits data. We then adopted the least-squares deconvolution (LSD; \cite{1997MNRAS.291..658D}, \cite{2010A&A...524A...5K}) to extract each line profile for each exposure. In this method, we regard a continuum of the observed spectrum as a convolution of a line profile and a series of delta functions, which can be obtained from a list of the absorption lines calculated using the Vienna Atomic Line Database (VALD; \cite{2000BaltA...9..590K}). After creating the line profiles, we made them smoother by averaging the five surrounding values for each data point.

We also analyzed the transit spectral dataset captured by EXPRES, whose resolution is also high ($R\sim 150,000$), mounted on the Lowell Discovery Telescope \citep{2012SPIE.8444E..19L} on UT 2020 August 2.
This dataset was used to detect its planetary shadow once in \citet{2021AJ....162..218C}. This dataset includes 41 spectra with an exposure time of 300 s and has already been shifted to a barycentric frame. These spectra have a signal-to-noise ratio per pixel of $\sim 30$.
We then corrected the continuum and telluric lines included in these fits data. The wavelength range adopted in this study is from 3990 \AA \ to 6540 \AA \, except for deep and wide lines, such as the Na D and H$\beta$ lines. The process of extracting each smooth line profile for each exposure is the same as that used to analyze the CARMENES dataset.

The line profile of the host star TOI-1518 was created by averaging line profiles during out-of-transit. To expose a dark track, called a planetary shadow, we subtracted the stellar line profile from each exposure line profile. We jointly fitted the stellar line profile and observed a planetary shadow to the models that adopted the MCMC using \texttt{EMCEE}.
The model of the stellar line profile is composed of an intrinsic stellar line profile and a broadening kernel due to stellar rotation and macro-turbulence \citep{2011ApJ...742...69H}. Considering the effect of stellar macro-turbulence, we modeled the planetary shadow by combining Equation (10) in \citet{2011ApJ...742...69H} and the equations in the Appendix of \citet{2020PASJ...72...19W}.

We derived the values of the following 20 parameters for MCMC fitting: 
impact parameter in 2020 $b_{2020}$ of each instrument, spin-orbit obliquity $\lambda$ of each instrument, the orbital period $P_{\mathrm{orb}}$, mid-transit time $T_{0}$, the radius ratio $R_{p}/R_{s}$, semi-major axis normalized by stellar radii $a/R_{s}$, two quadratic limb darkening coefficients $u_1$ and $u_2$ of CARMENES and EXPRES, stellar rotational velocity $V\sin i_s$, macro-turbulence velocity $v_{\mathrm{mac}}$, FWHM of intrinsic stellar line profile $v_{\mathrm{FWHM}}$, radial velocity of the planetary system $\gamma$, jitter terms for the stellar line profiles $\sigma_{\mathrm{jit, ste}}$ of CARMENES and EXPRES, and jitter terms for the planetary shadow $\sigma_{\mathrm{jit, ps}}$ of CARMENES and EXPRES. 
Here, we estimated the intrinsic stellar line profile as a Gaussian profile. 
In the MCMC fitting, we set the logarithm of the probability for each dataset of the stellar profile and planetary shadow to
\begin{eqnarray}
\label{post_tr}
\ln L_{\mathrm{prob}} =&-&\frac{1}{2} \sum_{i} \left[\frac{\left(O_{i}-C_{i}\right)^2}{\sigma_i^2+\sigma_{\mathrm{jit}}^2}+ n \ln \left\{2 \pi \left(\sigma_i^2+\sigma_{\mathrm{jit}}^2\right)\right\} \right] \nonumber \\
&-&\frac{1}{2}\sum_{j} \left(\frac{p_{j}-\mu_{j}}{s_j}\right)^2.
\end{eqnarray}
The first term in Equation (\ref{post_tr}) shows the logarithm of the likelihood, where $O_i$ is the data, $C_i$ is the model, and $\delta_i$ is the error of the $i$th data point. The second term represents the Gaussian priors; where $p_j$ is the parameter value, $\mu_j$ is the center value of the Gaussian prior, and $s_j$ is the uncertainty of the Gaussian prior.
We ran 20,000 steps, cut off the first 10,000 steps as burn-in, and iterated this set 48 times. Figure \ref{MCMC_SPEC} in the Appendix \rev{1} shows the posterior distributions.

We also derived these values using a bootstrap analysis for checking systematic errors. In this technique, we first compute the residuals by subtracting the best-fit model from the maximization of Equation \ref{post_tr} for the line profile data and Doppler tomographic data.
We then randomly shuffled the residuals with their errors and created new line profile data for the out-of-transit and Doppler tomographic datasets by adding the residuals and the best-fitting model.
A total of 200 fake datasets were created. We then executed MCMC fitting by running 4,000 steps, excluding the first 3,000 steps, and iterating 40 times for each mimic dataset. Figure \ref{MCMC_SPEC_BS} in the Appendix shows the distributions of the optimum values.
The systematic errors are negligible because the posterior distributions from the MCMC and the bootstrap methods are comparable. Consequently, we adopted the results from the MCMC method.

\begin{table*}
  \tbl{Measured parameters of TOI-1518b from CARMENES and EXPRES. In the column of "Priorfor Fitting", we describe uniform priors as $\mathcal{U}$(lower bound, upper bound) and normal priors as $\mathcal{N}$(mean, standard deviation).}{%
  \begin{tabular}{lccc}
      \hline
      Parameter & MCMC (Adopted) & Bootstrap& Prior for Fitting\\ 
      \hline
      Impact Parameter in 2020 from CARMENES $b_{2020, \mathrm{C}}$ & $0.8953 ^{+0.0050}_{-0.0048}$ &  $0.8954^{+0.0051}_{-0.0047}$ & $\mathcal{U}(0,1)$ \\
      Spin-orbit Obliquity from CARMENES $\lambda_{\mathrm{C}}$ (deg) & $-118.96 ^{+0.68}_{-0.65}$ & $-118.93^{+0.66}_{-0.65}$& $\mathcal{U}(-180,180)$ \\
      Impact Parameter in 2020 from EXPRES $b_{2020, \mathrm{E}}$ &$0.8574 ^{+0.0051}_{-0.0043}$ &  $0.8572^{+0.0049}_{-0.0043}$  & $\mathcal{U}(0,1)$ \\
      Spin-orbit Obliquity from EXPRES $\lambda_{\mathrm{E}}$ (deg)  &$-117.24 ^{+0.64}_{-0.69}$ & $-117.19^{+0.62}_{-0.70}$ & $\mathcal{U}(-180,180)$ \\
      Orbital Period $P_{\mathrm{orb}}$ (days)& $1.90261178 ^{+0.00000018}_{-0.00000019}$ & $1.90261178 \pm 0.00000019$    & $\mathcal{N}(1.90261178,0.00000019)$\footnotemark[$\ddag$]\\
      Mid-transit Time $T_{0}$ (BJD$_{\mathrm{TDB}}$)& $2459064.83065 \pm0.00010$ &$2459064.83066 \pm0.00010$  & $\mathcal{N}(2459064.83063,0.00010)$\footnotemark[$\ddag$]\footnotemark[$\S$]\\
      Radius Ratio $R_{p}/R_{s}$      & $0.10043\pm0.00019$  & $0.10043\pm0.00019$ & $\mathcal{N}(0.10037,0.00020)$\footnotemark[$\ddag$]\\
      Scaled Semi-major Axis $a/R_{s}$  & $4.171^{+0.015}_{-0.014}$ & $4.171\pm 0.015$   & $\mathcal{N}(4.167,0.014)$\footnotemark[$\ddag$] \\
      Limb-darkening Coefficient $u_1$ for CARMENES $u_{1,\mathrm{C}}$            & $0.4216 ^{+0.0021}_{-0.0022}$& $0.4215 \pm 0.0021$             & $\mathcal{N}(0.4226,0.0021)$\footnotemark[$*$] \\
      Limb-darkening Coefficient $u_2$ for CARMENES $u_{2,\mathrm{C}}$            & $0.1617 ^{+0.0031}_{-0.0034}$& $0.1617 \pm 0.0032$  & $\mathcal{N}(0.1628,0.0032)$\footnotemark[$*$] \\
      Limb-darkening Coefficient $u_1$ for EXPRES $u_{1,\mathrm{E}}$            & $0.5321 ^{+0.0027}_{-0.0028}$& $0.5320 ^{+0.0028}_{-0.0027}$   & $\mathcal{N}(0.5314,0.0028)$\footnotemark[$*$] \\
      Limb-darkening Coefficient $u_2$ for EXPRES $u_{2,\mathrm{E}}$            & $0.1672 \pm 0.0038$ & $0.1670 \pm 0.0038$  & $\mathcal{N}(0.1665,0.0038)$\footnotemark[$*$] \\
      Apparent Stellar Rotational Velocity $V\sin i_s$ (km s$^{-1}$)& $76.624 ^{+0.051}_{-0.052}$& $76.621 \pm 0.051$ & $\mathcal{U}(0,200)$ \\
      Macro-turbulence Velocity $v_{\mathrm{mac}}$ (km s$^{-1}$) & $13.90 ^{+0.46}_{-0.50}$& $13.91 ^{+0.47}_{-0.50}$ & $\mathcal{U}(0,50)$ \\
      FWHM of Gaussian Line Profile $v_{\mathrm{FWHM}}$ (km s$^{-1}$)  & $7.12 ^{+0.36}_{-0.35}$& $7.12 ^{+0.37}_{-0.36}$ & $\mathcal{N}(3.5,1)$\footnotemark[$\dag$] \\
      Radial Velocity of System $\gamma$ (km s$^{-1}$)  & $-12.671 ^{+0.038}_{-0.039}$& $-12.668 ^{+0.039}_{-0.038}$ & $\mathcal{U}(-30,30)$ \\
      Jitter for Stellar Line Profile of CARMENES $\sigma_{\mathrm{jit, ste, C}}$        & $0.00651 ^{+0.00052}_{-0.00047}$& $0.00651 ^{+0.00051}_{-0.00047}$ & $\mathcal{U}(0,1)$ \\
      Jitter for Planetary Shadow of CARMENES $\sigma_{\mathrm{jit, ps, C}}$       & $< 0.0011\ (3\sigma)$& $< 0.0011\ (3\sigma)$ & $\mathcal{U}(0,1)$ \\
      Jitter for Stellar Line Profile of EXPRES $\sigma_{\mathrm{jit, ste, E}}$     & $0.00483 ^{+0.00042}_{-0.00041}$& $0.00484 ^{+0.00044}_{-0.00041}$ & $\mathcal{U}(0,1)$ \\
      Jitter for Planetary Shadow of EXPRES $\sigma_{\mathrm{jit, ps, E}}$    & $< 0.0021\ (3\sigma)$& $< 0.0021\ (3\sigma)$ & $\mathcal{U}(0,1)$ \\
      \hline
    \end{tabular}}\label{CARresult}
\begin{tabnote}
\footnotemark[$*$] Median and standard deviation of these coefficients are calculated by \texttt{PyLDTk} \citep{Husser2013, Parviainen2015_b} using the values of $T_{\mathrm{eff}}$, $\log g$ and [Fe/H] from \citet{2021AJ....162..218C}.\\ 
\footnotemark[$\dag$]  To set the median and standard deviation of this parameter, we referred this value from the typical range of Gaussian dispersion of spectral lines from Table 1 in \citet{2011ApJ...742...69H}.\\
\footnotemark[$\ddag$] Median and standard deviation of these parameters are from the best values and the uncertainties of those derived in Section 2.1. \\
\footnotemark[$\S$] This mid-transit time is during the transit observation by EXPRES.
\\ 
\end{tabnote}
\end{table*}

\begin{figure*}[htbp]
 \begin{minipage}[b]{0.45\linewidth}
    \centering
    \hspace*{-7mm}
    \includegraphics[keepaspectratio, scale=0.4]{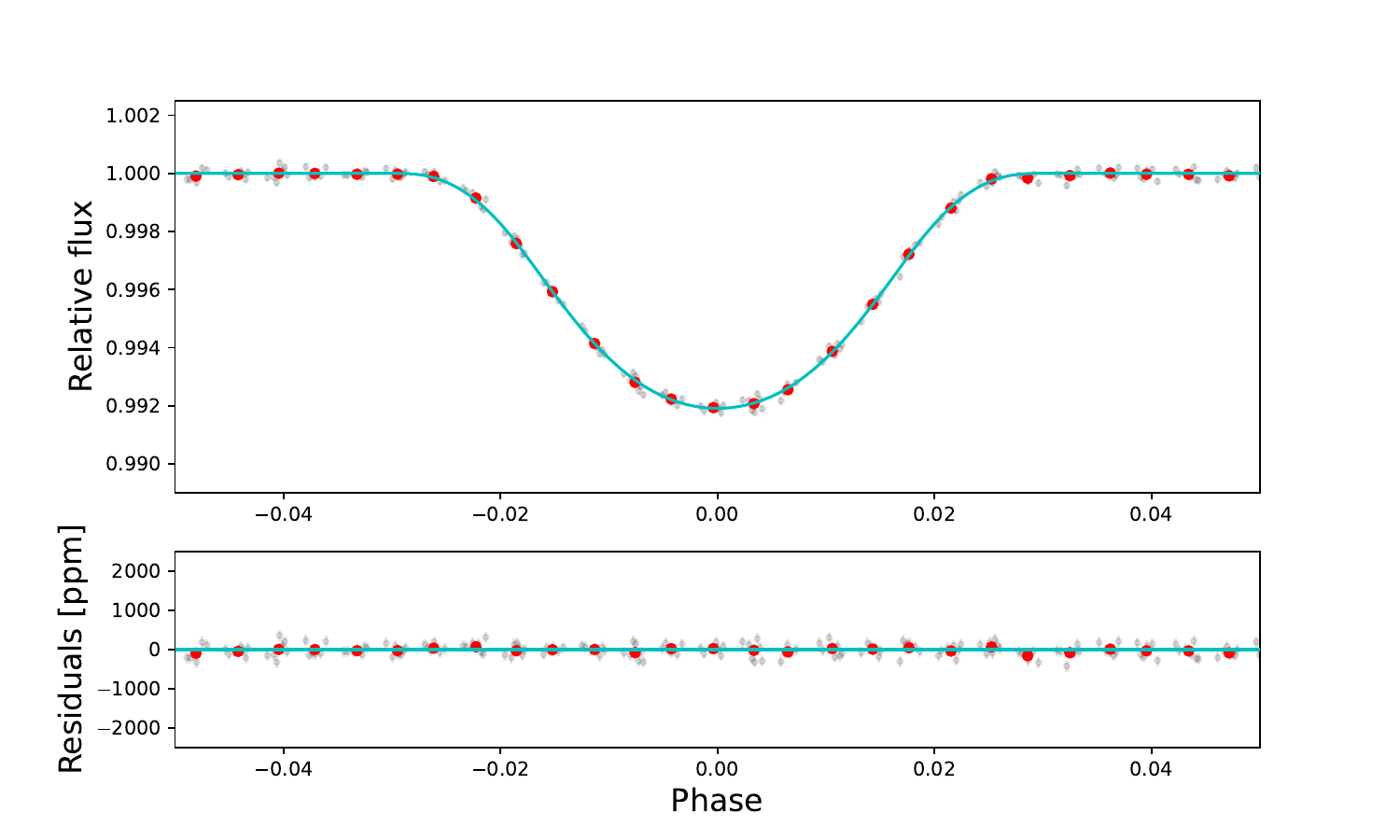}
  \end{minipage}
  \begin{minipage}[b]{0.7\linewidth}
    \centering
    \includegraphics[keepaspectratio, scale=0.4]{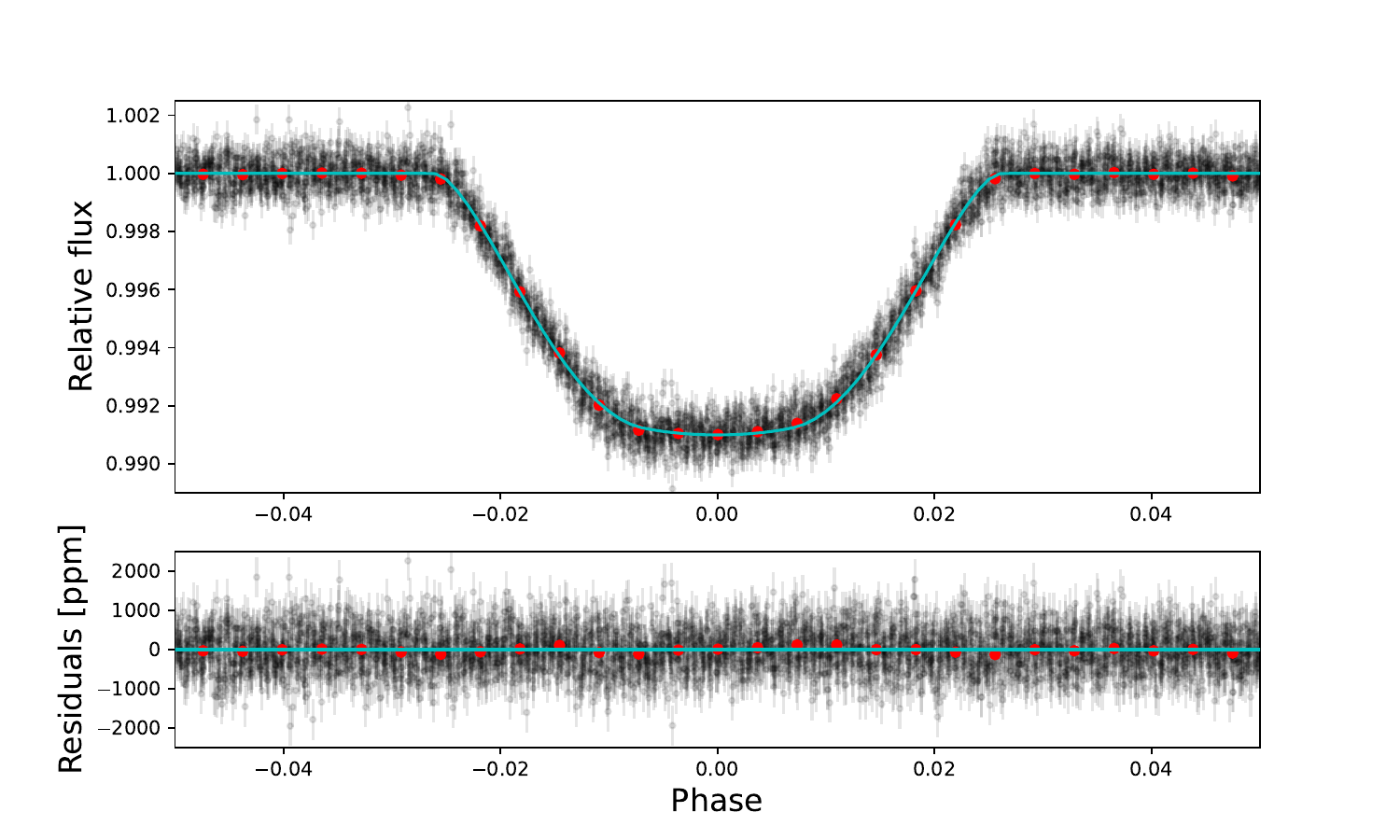}
  \end{minipage}
\caption{Phase-folded light curves of TOI-1518b in 2019 (left panel) and 2022 (right panel). These light curves are subtracted using the Gaussian process. The gray and red points show the observed data and the 10-minute binned data, respectively. The cyan lines show the best-fit light curve models. The bottom panels show the residuals between the observed data and model data.} \label{lcph}
\end{figure*}

\begin{figure*}[htbp]
 \begin{minipage}[b]{0.45\linewidth}
    \centering
    \hspace*{-7mm}
    \includegraphics[keepaspectratio, scale=0.4]{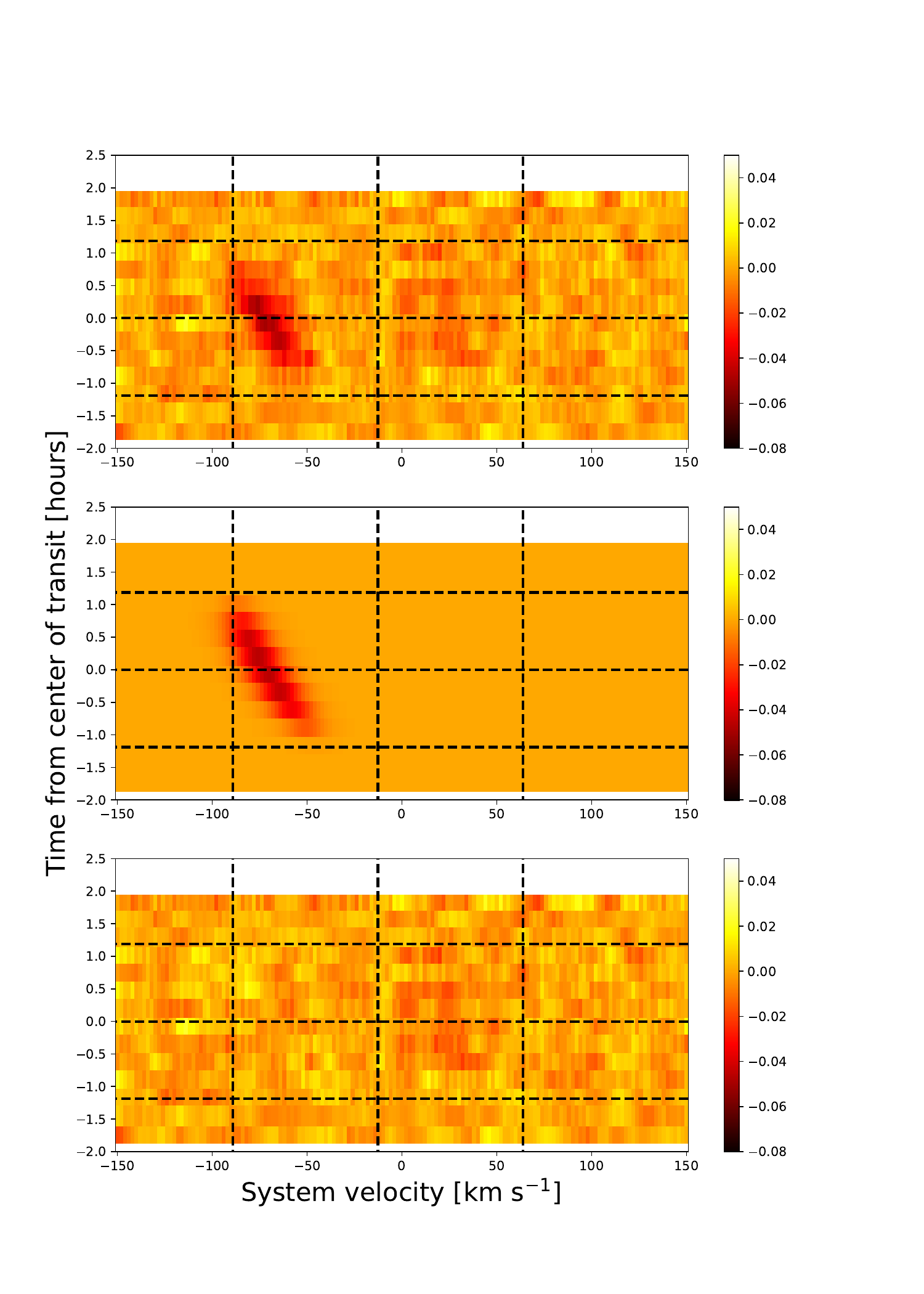}
  \end{minipage}
  \begin{minipage}[b]{0.7\linewidth}
    \centering
    \includegraphics[keepaspectratio, scale=0.4]{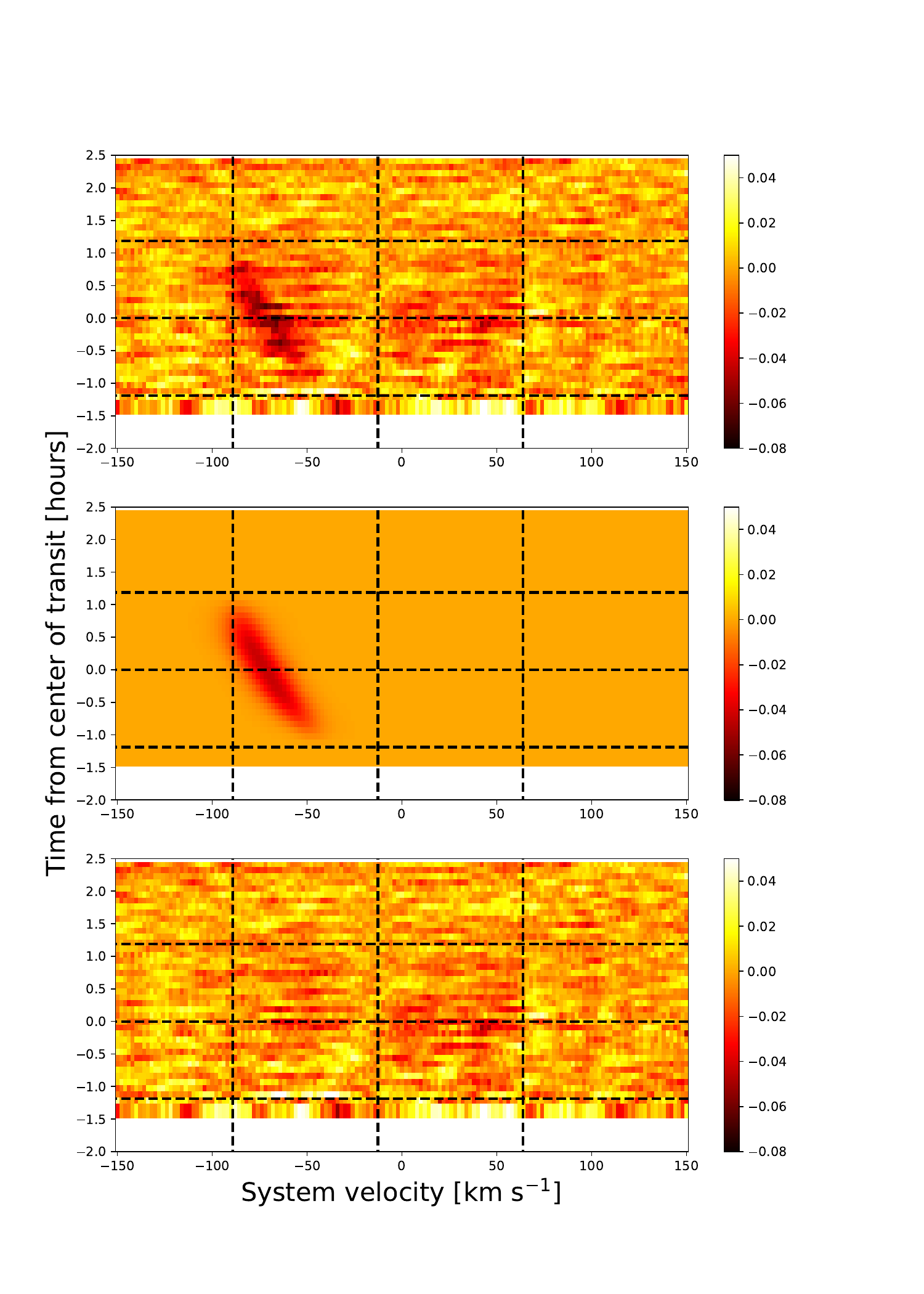}
  \end{minipage}
\caption{Doppler tomographic datasets of TOI-1518b from CARMENES (left) and EXPRES (right). Top panel: Observed residuals of line profile series from CARMENES. Middle panel: Model of planetary shadows using best-fit values via MCMC. Bottom panel: Difference between the top and the middle panels.} \label{resi}
\end{figure*}

\begin{figure*}[htbp]
 \begin{minipage}[b]{0.5\linewidth}
    \centering
    \hspace*{-7mm}
    \includegraphics[keepaspectratio, scale=0.5]{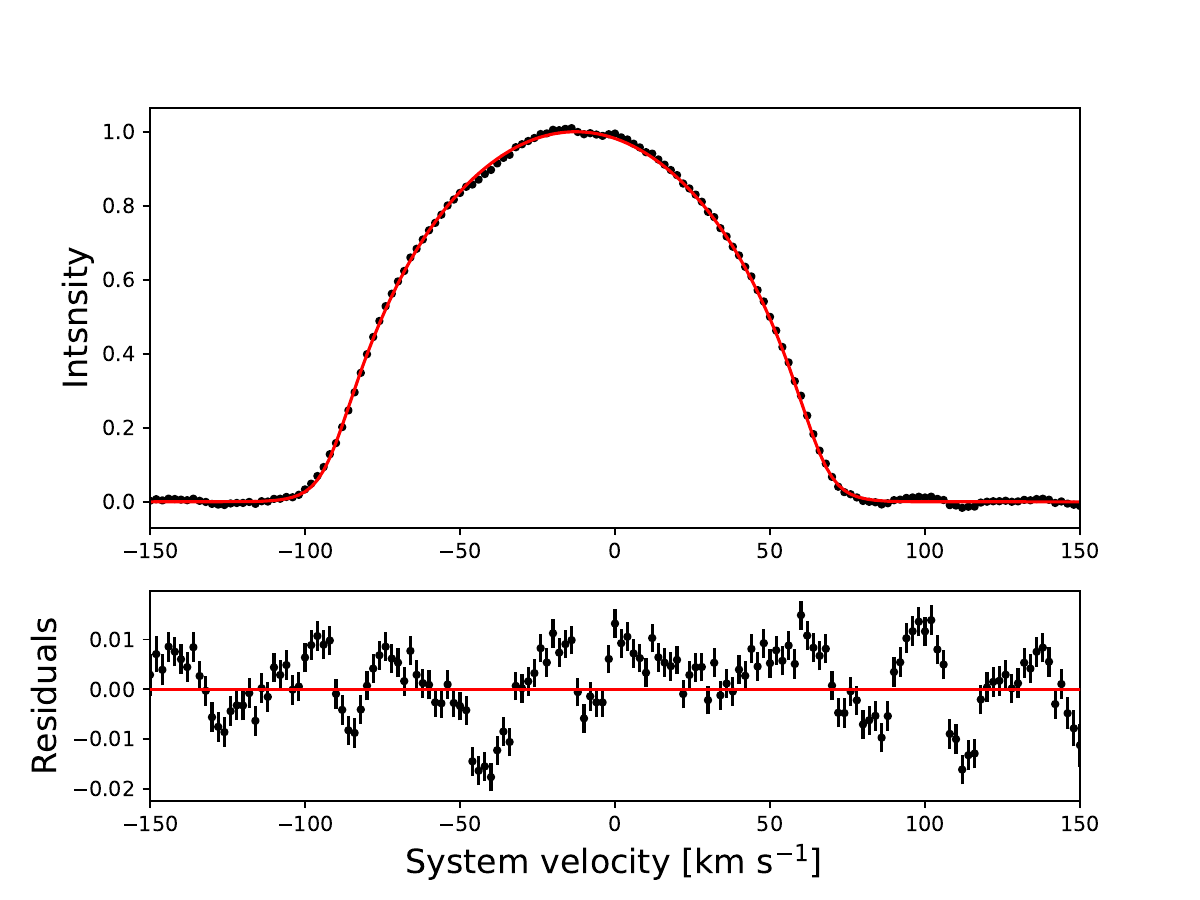}
  \end{minipage}
  \begin{minipage}[b]{0.5\linewidth}
    \centering
    \includegraphics[keepaspectratio, scale=0.5]{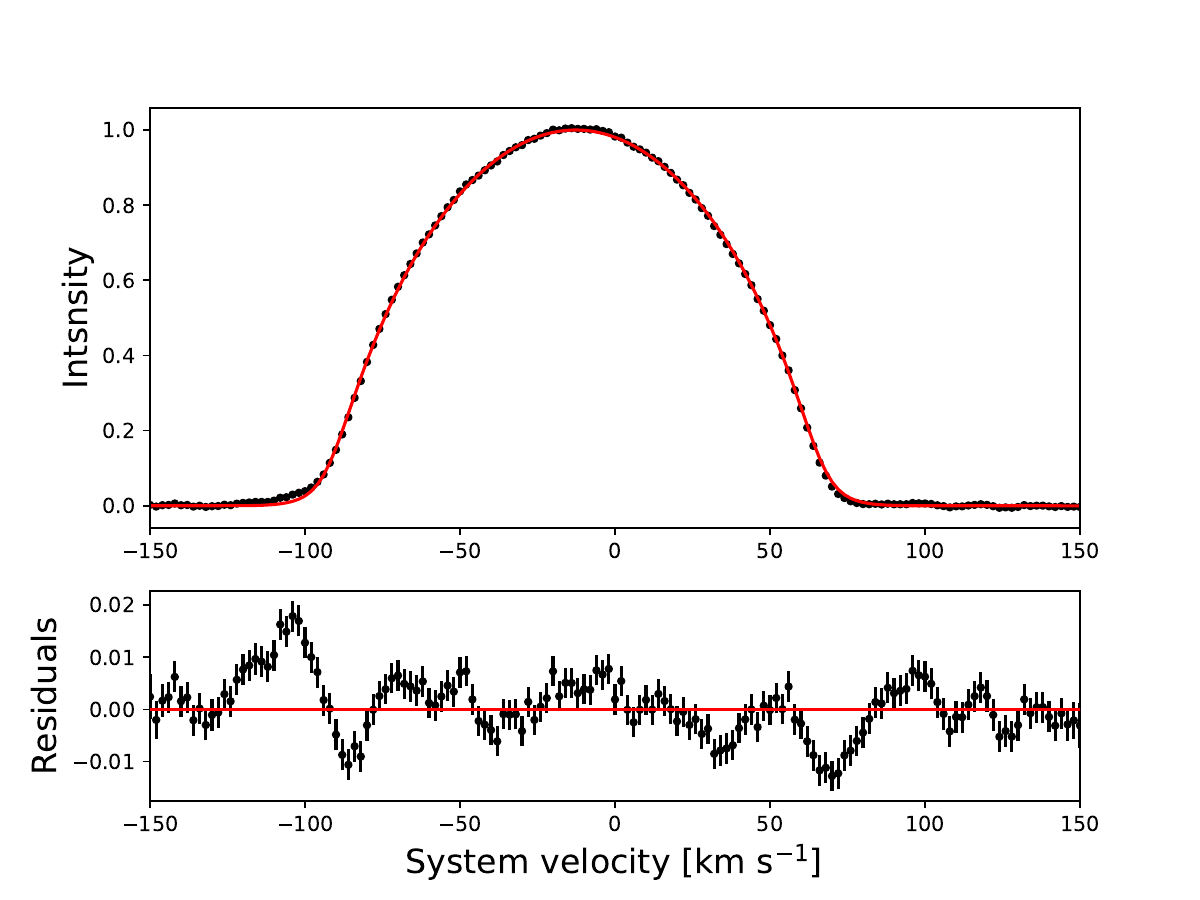}
  \end{minipage}
\caption{Averaged line profiles of TOI-1518 during out-of-transit from CARMENES (left, averaged 5 exposures) and EXPRES (right, averaged 15 exposures). 
The black dots are the observed data, and the red line shows the line profile model. The bottom panels show the residuals between the observed and model data. The different number of exposures should be a cause for their equivalent error bars although their signal-to-noises differ.} \label{lp_resi}
\end{figure*}

\section{Result and Discussion}
Figure \ref{lcph} shows the phase-folded TESS light curves and the best-fitting light curve models. Figure \ref{resi} displays the line profile residuals and the best-fitted models. Figure \ref{lp_resi} shows the averaged line profiles during out-of-transit and best-fitted line-profile models.
Tables \ref{TESSresult} and \ref{CARresult} list the best values and priors for the MCMC fitting from the transit photometric and spectral observations, respectively. Our values of $P_{\mathrm{orb}}$, $T_{0}$ and $R_p/R_s$ are consistent with those of \citet{2021AJ....162..218C} within $\sim 1 \sigma$ ($P_{\mathrm{orb}}=1.902603 \pm 0.000011$ days, $T_{0}=2458787.049255 \pm 0.000094$ BJD$_{\mathrm{TDB}}$
\footnote{From their values of $P_{\mathrm{orb}}$ and $T_{0}$, forecasted $T_{0}$ during the first transit in the TESS observation and the transit observation by EXPRES are $T_{0}=2458766.12062 \pm 0.00015$ BJD$_{\mathrm{TDB}}$ and $T_{0}=2459064.8293 \pm 0.0016$ BJD$_{\mathrm{TDB}}$, respectively.}
, $R_p/R_s=0.0988^{+0.0015}_{-0.0012}$ in \cite{2021AJ....162..218C}).  However, our values of $\lambda$ from EXPRES, $b$ in 2019 and $a/R_s$ and differ from those in \citet{2021AJ....162..218C} by $\sim 2 \sigma$ ($\lambda_{E}=\timeform{-119D66}^{+0.98}_{-0.93}$, $b_{2019}=0.9036^{+0.0061}_{-0.0053}$, $a/R_s=4.291^{+0.057}_{-0.061}$ in \cite{2021AJ....162..218C}).
In Table \ref{CARresult}, the values and uncertainties from the bootstrap are comparable to those from MCMC fitting. Here, we adopted the values obtained via the MCMC as the measured values.

\begin{figure}[htbp]
 \begin{center}
  \includegraphics[width=\linewidth]{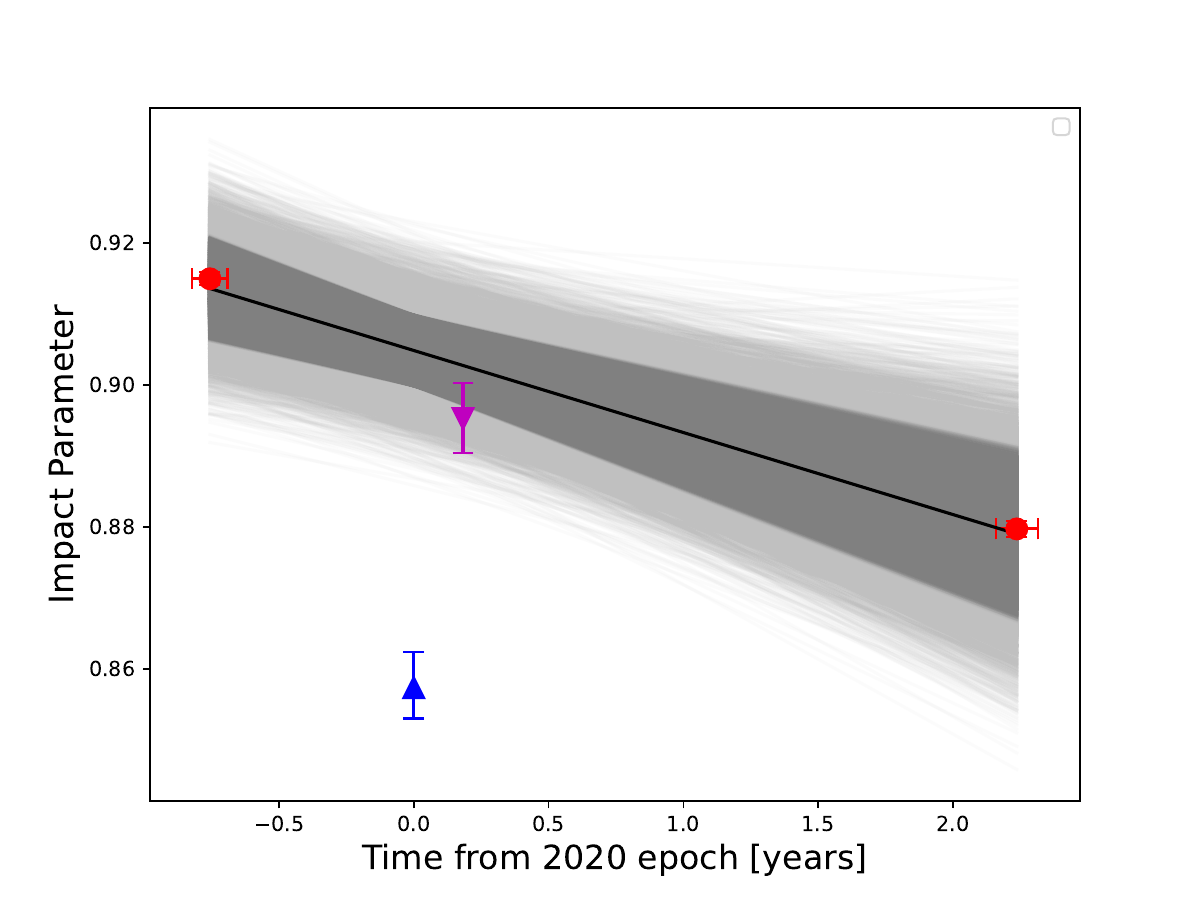} 
 \end{center}
 \begin{center}
  \includegraphics[width=\linewidth]{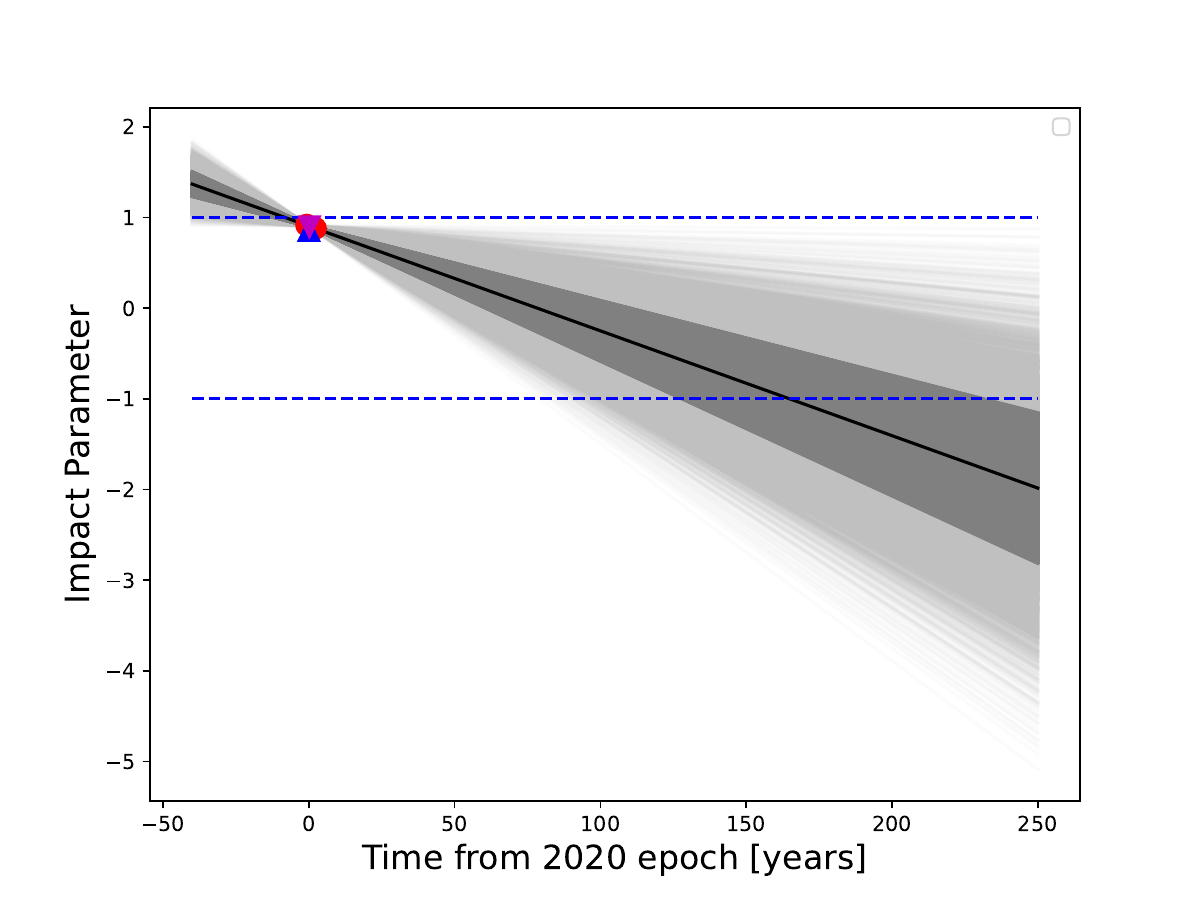} 
 \end{center}
\caption{Change in $b$ of TOI-1518b for short term (top) and long term (bottom). The red circles are values from TESS, the magenta inverted triangle is a value from CARMENES, the blue triangle is a value from EXPRES, and the black solid lines show the best-fit model of the nodal precession. The two blue-dashed lines show the edges of the stellar disc of TOI-1518b. The dark gray and light gray lines represent model likelihoods within the $1 \sigma$ and $3 \sigma$ confidence, respectively.}\label{prec}
\end{figure}

The values of $b$ in 2020 from CARMENES and EXPRES differ by $5\sigma$ in both analyses. Additionally, Figure \ref{prec} shows that the measured $b$ from CARMENES agrees with that from the precession model, whereas that from EXPRES disagrees with the model.
Concerning the dominant stellar rotational angular moment and orbital angular moment in this system, the impact parameter should not change immediately over several months. 
\rev{Observed planetary shadow from EXPRES in Figure \ref{resi} is less continual than that from CARMENES because the signal-to-noise of EXPRES is low. This would be the cause of the discrepancies.}
This may imply that the precession of TOI-1518b has a short-term variation due to other factors; \rev{a resonant normal mode of the host star driven by tidal excitation is one of the possible causes of the variation \citep{2023MNRAS.522.1968A}.}

%or that one of the datasets is inaccurate.

From the values calculated in Section 2, we derived the change in the impact parameter of TOI-1518b using weighted least squares. The value of $b$ is decreasing with $db/dt = -0.0116 \pm 0.0036 \mathrm{yr}^{-1}$. Figure \ref{prec} shows the changes in $b$ for TOI-1518b.
If $db/dt$ of TOI-1518\,b is constant while its transit trajectory is on the stellar disk, this transit would have begun at $2003^{+4}_{-7}$ CE and be going to end at $2194^{+70}_{-39}$ CE.

We derived an equation to estimate the nodal precession speed from $db / dt$.
Given the short orbital period of a planet and the rapid rotation of its host star, we can consider the stellar rotational vector as a steady vector. In this case, $b$ and $\lambda$ \rev{are expressed} as follows.
\begin{equation}
\label{bc}
b(t) =\frac{a}{R_s}(\cos \psi \cos i_s + \sin \psi \sin i_s \cos \theta(t))
\end{equation}
and
\begin{equation}
\label{lamc}
\tan \lambda(t) = \frac{\sin \psi \sin \theta(t)}{\sin \psi \cos i_s \cos \theta(t) - \cos \psi \sin i_s},
\end{equation}
where $\psi$ and $\theta(t)$ are the real spin-orbit obliquity and the nodal angle, respectively \citep{2022MNRAS.512.4404W}. \rev{The derivation of Equations (\ref{bc}) and (\ref{lamc}) is written in the Appendix 2.}
From \citet{2011Ap&SS.331..485I}, $\psi$ can be described as
\begin{equation}
\label{psi}
\cos \psi = \frac{b(t)R_s \cos i_s}{a} + \sin i_s \cos \lambda(t)  \sqrt{1-\left(\frac{b(t)R_s}{a}\right)^2}.
\end{equation}
Using Equations (\ref{bc}), (\ref{lamc}), and (\ref{psi}), the nodal precession speed $d \theta /dt$ can be expressed as
\begin{equation}
\label{dthdt}
\frac{d\theta}{dt} =\frac{1}{\sin i_s \sin \lambda \sqrt{\left(\frac{a}{R_s}\right)^2 - b^2}}\frac{db}{dt},
\end{equation}
where we assume that $\psi$ and $i_s$ are constants and $\theta$ is a time-variable. 
\citet{2013ApJ...774...53B} also give the expression of $d\theta / dt$ for a planet in a circular orbit as
\begin{equation}
\label{prec_e}
\frac{d\theta}{dt} = -\frac{3 \pi J_2 R_s^2 \cos \psi}{P_{\mathrm{orb}}a^2},
\end{equation}
where $J_2$ is the stellar quadrupole moment.
 
We impose a-priori constraints on $i_s$ \citep{2011Ap&SS.331..485I},
\begin{equation}
\label{ilim}
\sin i_s > V_s \sqrt{\frac{R_s}{GM_s}},
\end{equation}
where $V_s (=V\sin i_s / \sin i_s)$ denotes the stellar rotation speed. This equation was derived based on the condition that the gravitational acceleration at the stellar surface should be greater than the centrifugal acceleration at the stellar equator. The value of $i_s$ for TOI-1518 is estimated to be $\timeform{9.71D} \leq i_s \leq \timeform{170.29D}$ using our value of $V \sin i_s$ for $V_s$ and the values of $R_s$ and $M_s$ from \citet{2021AJ....162..218C}. Then, we set the range of $\psi$ to $\timeform{81.84D} \leq \psi \leq \timeform{121.98D}$ from Equation (\ref{psi}) using our values of $b$, $a/R_s$, and $\lambda$ within an $1\sigma$-confidence level. However, considering $db/dt<0$, $\sin \lambda<0$ and $\sin i_s >0$, $d\theta / dt$ must be positive from Equation (\ref{dthdt}), and $\psi$ must be greater than $\timeform{90D}$ from Equation (\ref{prec_e}). Thus, the possible ranges of $\psi$ and $i_s$ should be $\timeform{90D} < \psi \leq \timeform{121.98D}$ and $\timeform{22.86D} \leq i_s \leq \timeform{170.28D}$, respectively. Therefore, we can set the limit of the nodal precession speed as $\timeform{0.13D} \mathrm{yr}^{-1} \leq d\theta / dt \leq \timeform{1.43D} \mathrm{yr}^{-1}$ from Equation (\ref{dthdt}), and the lower limit of the stellar quadrupole moment of TOI-1518 as $J_{2, \mathrm{min}}=4.41 \times 10^{-5}$ from equation (\ref{prec}). 
The minimum value of $J_2$ for TOI-1518 indicates that the shape of TOI-1518 is more oblate than that of the Sun ($J_{2,\odot}\sim 2 \times 10^{-7}$; \cite{2001A&A...377..688R}). TOI-1518 has a flattened shape like other rapidly-rotating hot stars such as Kepler-13A ($J_{2}=(6.1\pm 0.3)\times 10^{-5}$; \cite{2015ApJ...805...28M}), WASP-33 ($J_{2}=(1.36^{+0.15}_{-0.12})\times 10^{-4}$; \cite{2022MNRAS.512.4404W}), and KELT-9 ($J_{2}=(3.26^{+0.93}_{-0.80})\times 10^{-4}$; \cite{2022ApJ...931..111S}). \rev{These values of $J_{2}$ also indicate that hot stars are more likely to redistribute their internal mass than the Sun.}

We calculated the Love number $k_2$ of TOI-1518, which is an index of the stellar rigidity.
From Equations (2) and (3) in \citet{2009ApJ...698.1778R}, $J_2$ can be expressed with as $k_2$ 
\begin{equation}
\label{J2other_2}
J_2=\frac{k_2 R_s^3}{3 a^3}\left( \frac{P_\mathrm{orb}^2}{P_\mathrm{spin}^2}+\frac{3M_p}{2M_s} \right),
\end{equation}
where $M_p$ is the planetary mass and $P_\mathrm{spin}(=2\pi R_s / V_s)$ is the stellar rotation period. 
$P_\mathrm{spin}$ varies $ 0.21\,\mathrm{days}<P_\mathrm{spin}<1.31\,\mathrm{days}$ within the range of $i_s$. Thus, the range of $P_\mathrm{spin}/P_\mathrm{orb}$ is $0.11<P_\mathrm{spin}/P_\mathrm{orb}<0.69$.
The upper limit of the mass of TOI-1518\,b is $M_p<2.3M_J$ \citep{2021AJ....162..218C}, which gives $M_p / M_s < 0.0014$.
The second term in Equation (\ref{J2other_2}) should be negligible for this system.

We made 10,000 samples by randomly selecting values of $P_\mathrm{orb}$, $a/R_s$, $\psi$, $i_s$ and $d\theta/dt$ within the $1\sigma$ ranges for $P_\mathrm{orb}$ and $a/R_s$, and within the certain ranges for $\psi$, $i_s$ and $d\theta/dt$ ($\timeform{90D} < \psi < \timeform{121.98D}$, $\timeform{22.86D} < i_s \leq \timeform{170.28D}$, $\timeform{0.13D} \mathrm{yr}^{-1} < d\theta / dt < \timeform{1.43D} \mathrm{yr}^{-1}$).
Using equations (\ref{prec_e}) and (\ref{J2other_2}), we determine the distribution of $k_2$ from 10,000 sample systems.
We obtained $\log{k_2}= -2.17\pm 0.33$ for the Love number of TOI-1518, which is smaller than that of a sun-like star ($\log k_2 \sim -1.52$; \cite{1995A&AS..114..549C}). This result is consistent with the model of the relationship between $M_s$ and $k_2$ in \citet{1995A&AS..114..549C}.
For reference, we calculated the $k_2$ values of other hot stars from Equation (\ref{J2other_2}). These calculated values are listed in Table \ref{k2k2}. The Love number of TOI-1518 is similar to that of the other hot stars. This implies that the interior of a hot star is stiffer and less susceptible to tidal deformation than that of a sun-like star.

\begin{table*}
  \tbl{Calculated $\log k_2$ of hot stars and referred parameters for equations (\ref{prec_e}) and (\ref{J2other_2}).}{%
  \begin{tabular}{l|cccc}
      \hline
      Star & TOI-1518 & WASP-33 & KELT-9 & Kepler-13A\\
      \hline
      \hline
      $\log k_2$ &$-2.17\pm 0.33$& $-2.14 \pm 0.11$ & $-2.16^{+0.26}_{-0.30}$& $-2.17^{+0.22}_{-0.20}$ \\
      \hline
      $J_2$ &$>4.41 \times 10^{-5}$\footnotemark[(i)]& $(1.36^{+0.15}_{-0.12})\times 10^{-4}$\footnotemark[(iii)]& $(3.26^{+0.93}_{-0.80})\times 10^{-4}$\footnotemark[(vii)]& $(6.1\pm 0.3)\times 10^{-5}$\footnotemark[(x)] \\
      $\psi$ (deg) &$90 < \psi < 121.98$\footnotemark[(i)]& $108.09^{+0.95}_{-0.97}$\footnotemark[(iii)]& $87^{+10}_{-11}$\footnotemark[(viii)]& $60\pm2$\footnotemark[(x)] \\
      $d\theta/dt$ (deg yr$^{-1}$)&$0.13 < d\theta / dt < 1.43 $\footnotemark[(i)]& $0.507^{+0.025}_{-0.022}$\footnotemark[(iii)]& $0.404^{+0.076}_{-0.074}$\footnotemark[(vii)]\footnotemark[$*$]& $0.240^{+0.037}_{-0.028}$\footnotemark[(x)]\footnotemark[$*$] \\
      $i_s$ (deg)&$22.86 \leq i_s < 170.28$\footnotemark[(i)]&  $58.3^{+4.6}_{-4.2}$\footnotemark[(iii)]& $139\pm7$\footnotemark[(vii)]& $81.8\pm0.2$\footnotemark[(x)] \\
      $V \sin i_s$ (km s$^{-1}$) &$76.624^{+0.051}_{-0.052}$\footnotemark[(i)]&  $86.63^{+0.37}_{-0.32}$\footnotemark[(iv)] & $111.4\pm 1.3$\footnotemark[(ix)]& $78\pm 15$\footnotemark[(x)] \\
      $R_s$ ($R_\odot$)&$1.950\pm 0.048$\footnotemark[(ii)]&  $1.444\pm 0.034$\footnotemark[(v)] & $2.362^{+0.075}_{-0.063}$\footnotemark[(ix)]& $1.74\pm 0.04$\footnotemark[(xi)] \\
      $a/R_s$ &$4.171^{+0.015}_{-0.014}$\footnotemark[(i)]&  $3.69\pm 0.01$\footnotemark[(v)] & $3.153\pm0.011$\footnotemark[(ix)]& $4.5007^{+0.0039}_{-0.0040}$\footnotemark[(xi)] \\
      $M_s$ ($M_\odot$)&$1.79\pm0.26$\footnotemark[(ii)]&  $1.495\pm 0.031$\footnotemark[(v)] & $2.52^{+0.25}_{-0.20}$\footnotemark[(ix)]& $1.72\pm0.10$\footnotemark[(xi)] \\
      $M_p$ ($M_J$) &$<2.3$\footnotemark[(ii)]&  $2.81\pm 0.53$\footnotemark[(vi)] & $2.88\pm0.84$\footnotemark[(ix)]& $9.28\pm 0.16$\footnotemark[(xi)] \\
      $P_\mathrm{orb}$ (days) & $1.90261178^{+0.00000018}_{-0.00000019}$\footnotemark[(i)]&  $1.2198675\pm 0.0000011$\footnotemark[(vi)] & $1.4811235\pm 0.0000011$ \footnotemark[(ix)]& $1.763588\pm0.000001$\footnotemark[(xi)] \\
      \hline
    \end{tabular}}\label{k2k2}
\begin{tabnote}
\footnotemark[(i)]This work,\footnotemark[(ii)]\citet{2022MNRAS.512.4404W},\footnotemark[(iii)]\citet{2021AJ....162..218C}, \footnotemark[(iv)]\citet{2015ApJ...810L..23J}, \footnotemark[(v)]\citet{2010MNRAS.407..507C}, \footnotemark[(vi)]\citet{2014A&A...561A..48V}, \footnotemark[(vii)]\citet{2022ApJ...931..111S}, \footnotemark[(viii)]\citet{2020AJ....160....4A},\footnotemark[(ix)]\citet{2017Natur.546..514G}, \footnotemark[(x)]\citet{2015ApJ...805...28M}, \footnotemark[(xi)]\citet{2015ApJ...804..150E}\\
\footnotemark[$*$] These values are derived from the nodal precession period presented in the earlier studies. \\
\end{tabnote}
\end{table*}

\section{Conclusion}
We investigated the nodal precession of TOI-1518\,b using transit photometric datasets from TESS and Doppler tomographic datasets from CARMENES and EXPRES and measured the change in its impact parameter $db/dt = -0.0116 \pm 0.0036\,\mathrm{yr}^{-1}$. TOI-1518\,b is the fourth planetary system in which the nodal precession is detected.
We estimate that the transit has started in $2003^{+4}_{-8}$ CE and will cease in $2194^{+70}_{-39}$ CE if $b$ changes linearly while the transit trajectory is on the stellar surface. This result suggests that despite the nodal precession of TOI-1518\,b, its transit is observable with the next-generation telescopes such as Ariel \citep{2018ExA....46..135T} and the Thirty Meter Telescope (TMT).
We calculated the minimum value of $J_2$ for TOI-1518 $J_{2,\mathrm{min}}=4.41\times 10^{-5}$, which indicates that TOI-1518 is more oblate \rev{and more prone to their internal mass redistribution} than the sun. We also derived the logarithm of $k_2$ for TOI-1518 $\log{k_2}= -2.17\pm 0.33$. TOI-1518 may have a stiffer interior and may be less susceptible to tidal effects than the sun-like star.

The TESS mission has discovered hot Jupiters around the B- and A-type stars.
The real spin-orbit obliquity $\psi$ of these systems is key to revealing the orbital evolution of a hot Jupiter. 
Measuring $k_2$ of hot stars and $\psi$ through nodal precession observations every few years \citep{2022MNRAS.512.4404W} would provide insights into the tidal evolution and orbital migration of a planetary system with a hot Jupiter around a hot star.

\begin{ack}
This work is partly supported by JSPS KAKENHI Grant Numbers JP18H05439, JP21K20376, and JST CREST Grant Number JPMJCR1761.
This paper includes data collected with the TESS mission, obtained from the MAST data archive at the Space Telescope Science Institute (STScI). Funding for the TESS mission is provided by the NASA Explorer Program. STScI is operated by the Association of Universities for Research in Astronomy, Inc., under NASA contract NAS 5–26555. Resources supporting this work were provided by the NASA High-End Computing (HEC) Program through the NASA Advanced Supercomputing (NAS) Division at Ames Research Center for the production of the SPOC data products. We acknowledge the use of TESS High Level Science Products (HLSP) produced by the Quick-Look Pipe-line (QLP) at the TESS Science Office at MIT, which are publicly available from the Mikulski Archive for Space Telescopes (MAST). This work used data from the EXtreme PREcision Spectrograph (EXPRES) that was designed and commissioned at Yale with financial support by the U.S. National Science Foundation under MRI-1429365 and ATI1509436 (PI D. Fischer). We gratefully acknowledge support for telescope time using EXPRES at the LDT from the Heising-Simons Foundation and an anonymous Yale donor.
CARMENES was funded by the Max-Planck-Gesellschaft (MPG), the Consejo Superior de Investigaciones Científicas (CSIC), the Ministerio de Economía y Competitividad (MINECO), and the European Regional Development Fund (ERDF) through projects FICTS-2011-02, ICTS-2017-07-CAHA-4, and CAHA16-CE-3978, and the members of the CARMENES Consortium (Max-Planck-Institut für Astronomie, Instituto de Astrofísica de Andalucía, Landessternwarte Königstuhl, Institut de Ciències de l’Espai, Institut für Astrophysik Göttingen, Universidad Complutense de Madrid, Thüringer Landessternwarte Tautenburg, Instituto de Astrofísica de Canarias, Hamburger Sternwarte, Centro de Astrobiología and Centro Astronómico Hispano-Alemán), with additional contributions by the MINECO, the Deutsche Forschungsgemeinschaft (DFG) through the Major Research Instrumentation Programme and Research Unit FOR2544 “Blue Planets around Red Stars”, the Klaus Tschira Stiftung, the states of Baden-Württemberg and Niedersachsen, and by the Junta de Andalucía.
We would like to thank E. Palle for advising us on the proposal for CARMENES and helping to revise the manuscript.
We would also like to thank S. H. C. Cabot for providing spectral data of EXPRES.
This work was partially supported by the Research Fund for Students (2020) of the Department of Astronomical Science, SOKENDAI (the Graduate University for Advanced Studies).
This work has made use of the Vienna Atomic Line Database (VALD) database, operated at Uppsala University, the Institute of Astronomy RAS in Moscow, and the University of Vienna.
We would like to thank Editage (www.editage.jp) for English language editing.
%This paper includes data collected with the TESS mission, obtained from the MAST data archive at the Space Telescope Science Institute (STScI). Funding for the TESS mission is provided by the NASA Explorer Program. STScI is operated by the Association of Universities for Research in Astronomy, Inc., under NASA contract NAS 5–26555.
\end{ack}

\appendix

\section{MCMC Results of Photometric and Doppler Tomographic Measurements}
\begin{figure*}[htbp]
 \centering
 \hspace*{0mm}
 \includegraphics[width=\linewidth]{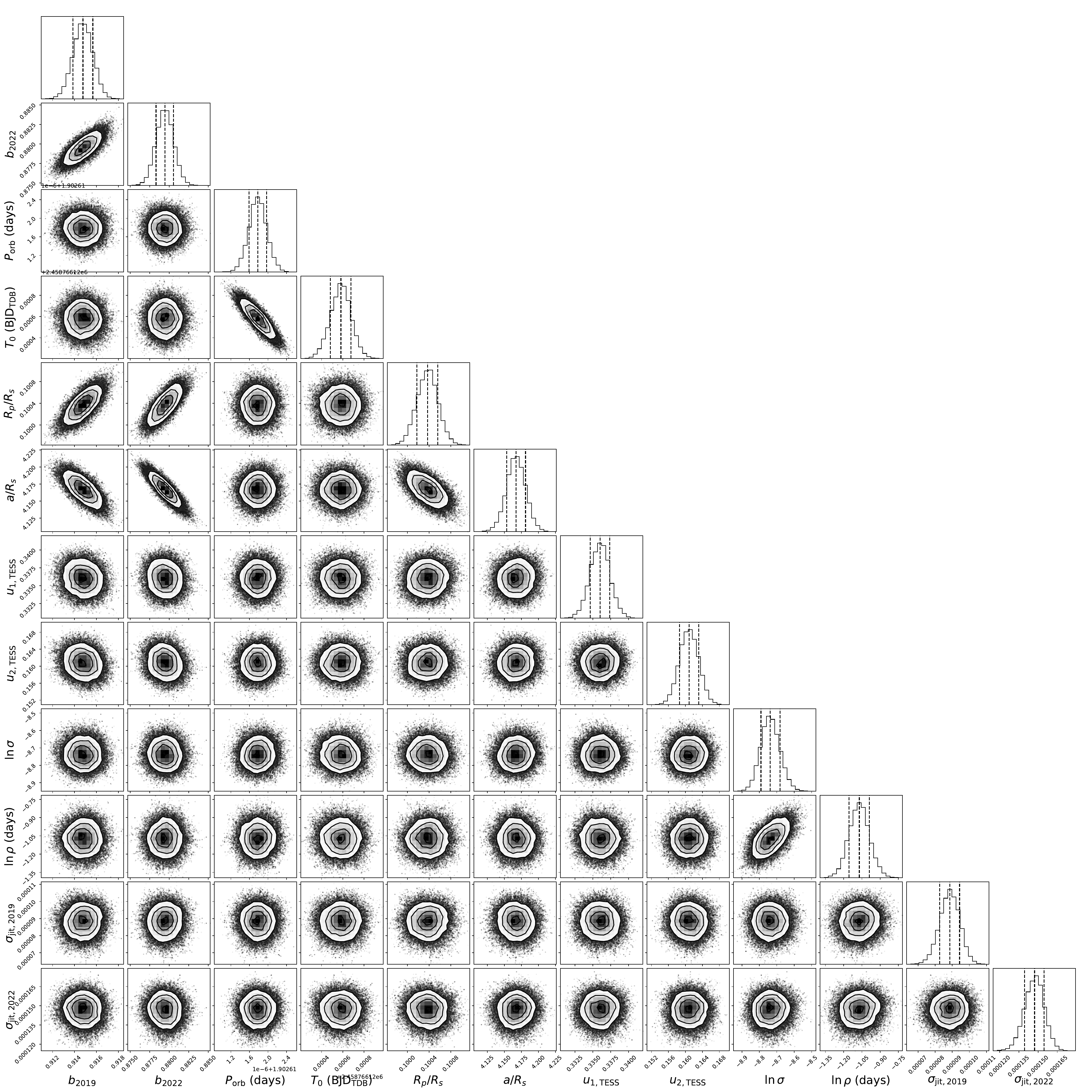}
\caption{Corner plots for the free parameters from the photometric datasets of TESS. We created these plots with corner.py \citep{corner}.} \label{MCMC_TESS}
\end{figure*}

\begin{figure*}[htbp]
 \centering
 \hspace*{0mm}
 \includegraphics[width=\linewidth]{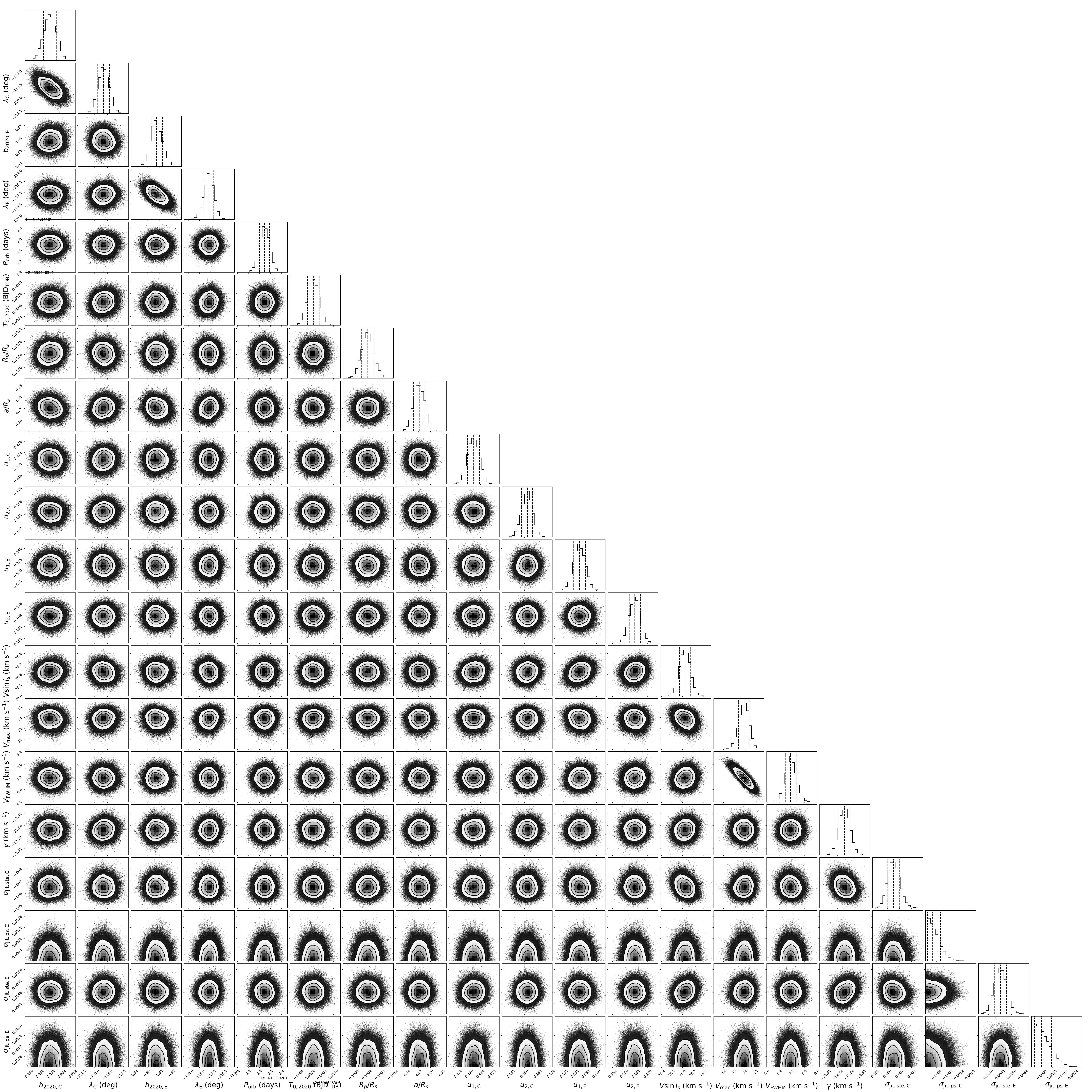}
\caption{Corner plots for the free parameters from the spectral datasets of CARMENES and EXPRES via MCMC method. We created these plots in the same way as Figure \ref{MCMC_TESS}.} \label{MCMC_SPEC}
\end{figure*}

\begin{figure*}[htbp]
 \centering
 \hspace*{0mm}
 \includegraphics[width=\linewidth]{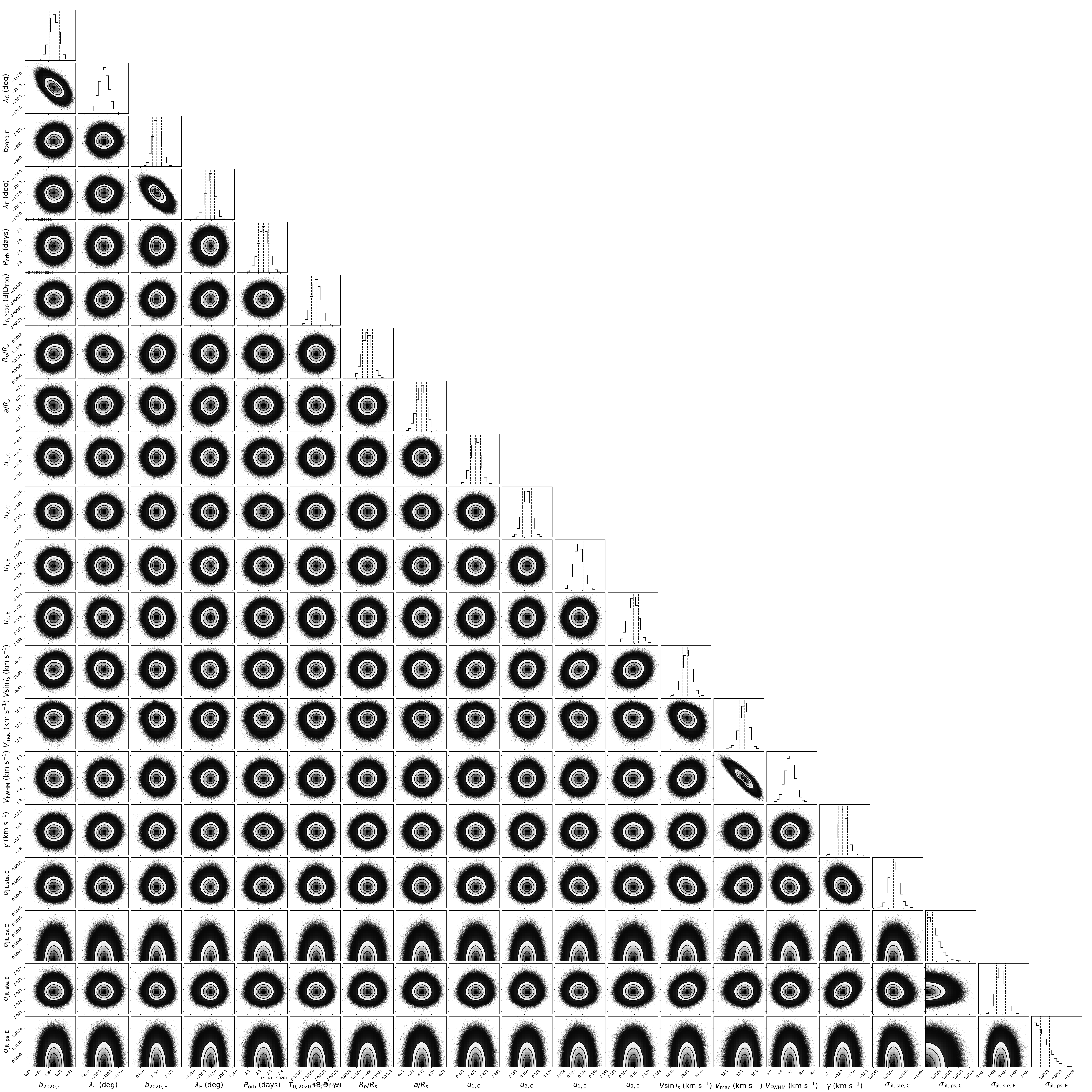}
\caption{Corner plots for the free parameters from the spectral datasets of CARMENES and EXPRES via bootstrap method. We created these plots in the same way as Figure \ref{MCMC_TESS}.} \label{MCMC_SPEC_BS}
\end{figure*}

In this appendix, we display the corner plots of posteriors after calculating by MCMC in Figures \ref{MCMC_TESS} and \ref{MCMC_SPEC}. We also show the corner plots of posteriors via bootstrap analysis in Figure \ref{MCMC_SPEC_BS}.

\section{\rev{Derivation of Changes of Impact Parameter and Projected Spin-orbit Obliquity}}
\begin{figure}[htbp]
% \hspace*{0mm}
\includegraphics[width=\linewidth]{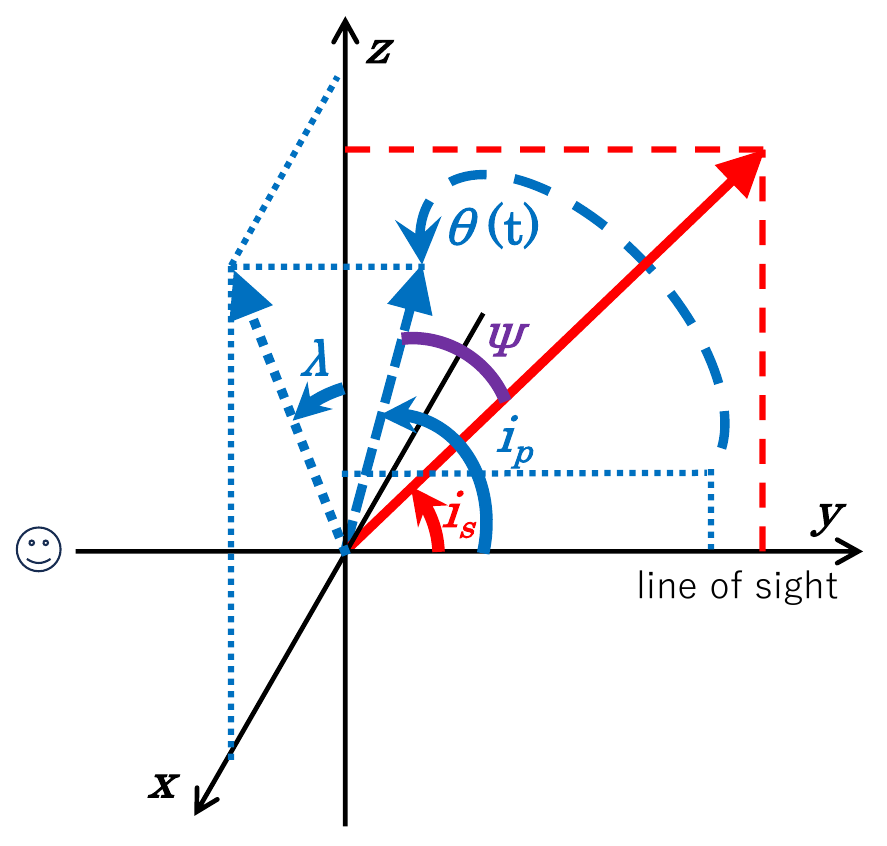}
\caption{\rev{Cartoon of planetary system. We set $y$ axis as the line of sight and $xz$ plane as the plane of sky. The solid red vector and the solid blue vector show the stellar rotational axis and the planetary orbital axis, respectively. We define $\theta (t)=0$ when the planetary orbital axis is on $yz$ plane.}} \label{cart}
\end{figure}

\rev{We define the angles $i_s$, $\theta (t)$, $\lambda$, $\psi$ and $i_p$ (the planetary inclination: $i_p=\arccos{(bR_s/a)}$) in Figure \ref{cart}. 
Considering the coordinate system with $z$ axis and stellar rotation axis aligned is rotated by $\timeform{90D} - i_s$ around $x$ axis, we can write the unit vector of the planetary orbital axis $\vec{k_p}$ as
\begin{eqnarray}
\label{kp1}
\vec{k_p}&=&
    \left(
    \begin{array}{ccc}
    1 & 0 & 0 \\
    0 & \sin{i_s} & \cos{i_s} \\
    0 & -\cos{i_s} & \sin{i_s}
    \end{array}
    \right)
    \left(
    \begin{array}{c}
    -\sin{\psi}\sin{\theta (t)} \\
    \sin{\psi}\cos{\theta (t)} \\
    \cos{\psi} 
    \end{array}
    \right)\nonumber \\ 
    &=&
    \left(
    \begin{array}{c}
    -\sin{\psi}\sin{\theta (t)} \\
    \cos{\psi}\cos{i_s}+\sin{\psi}\sin{i_s}\cos{\theta (t)} \\
    \cos{\psi}\sin{i_s}-\sin{\psi}\cos{i_s}\cos{\theta (t)}
    \end{array}
    \right).
\end{eqnarray}
$\vec{k_p}$ also can be expressed as
\begin{equation}
\label{kp2}
\vec{k_p}=
    \left(
    \begin{array}{c}
    \sin{i_p}\sin{\lambda} \\
    \cos{i_p} \\
    \sin{i_p}\cos{\lambda}
    \end{array}
    \right),
\end{equation}
which is also described in \citet{2016MNRAS.455..207I} as Equations (21), (22) and (23).
From $y$ components of Equations (\ref{kp1}) and (\ref{kp2}), we can derive the change in $b$ as
\begin{eqnarray}
\label{Abt}
b(t) &=& \frac{a}{R_s}\cos{i_p} \nonumber \\ 
     &=&\frac{a}{R_s}(\cos \psi \cos i_s + \sin \psi \sin i_s \cos \theta(t)).
\end{eqnarray}
On the other hand, dividing $x$ component of $\vec{k_p}$ by its $z$ component, we can obtain the change in $\lambda$ as
\begin{equation}
\label{Alt}
\tan \lambda(t) = \frac{\sin \psi \sin \theta(t)}{\sin \psi \cos i_s \cos \theta(t) - \cos \psi \sin i_s}.
\end{equation}
}

\bibliographystyle{apj}
\bibliography{reference}

\end{document}